\documentclass[aps,pra, onecolumn,unsortedaddress]{revtex4-1}
\usepackage{graphicx}
 \usepackage{amsmath}
 \usepackage{footnote}
 \usepackage{multirow}
 \usepackage{url}

\begin{document}
\title{Estimation of antihydrogen properties in experiments with small signal deficit}

\author{B. Radics}
\email{bradics@phys.ethz.ch}
\affiliation{ETH Z\"urich, Institute for Particle Physics and Astrophysics, CH-8093 Z\"urich, Switzerland}

\begin{abstract}
For a class of precision CPT-invariance test measurements using antihydrogen, a deficit in the data indicates the presence of the signal. The construction of classical confidence intervals for the properties of the antiatoms from measurements may pose a challenge due to the limited statistics experimentally available. We use the Feldman-Cousins \cite{FeldmanCousins98} method to estimate model parameters for such a low count rate measurement. First, we construct confidence intervals for the Poisson process with a known background and an unknown signal deficit. Then the generalized Monte Carlo version of the method is applied to the use case of the hyperfine transition frequency measurement of the ground-state antihydrogen atom, where the expected double-dip resonance line shape and the mean background is assumed to be known. We find that confidence intervals of the antihydrogen properties could be obtained already from low statistics data. We also discuss how the method may be extended to allow estimation of additional model parameters.
\end{abstract}

\maketitle

%----------------------------------------------------------------------------------------
%	ARTICLE CONTENTS
%----------------------------------------------------------------------------------------

\section{Introduction}
It is often the case in physics that a new phenomenon manifests itself as a small (positive) signal excess on top
of a well-understood background. In addition, searches for rare processes might require analysis of experimental data with few counts. The construction of classical (frequentist)
confidence intervals for such cases has been studied in great detail, and the most frequently used model processes, Poisson and Gaussian, have been
discussed by Feldman and Cousins \cite{FeldmanCousins98} among others. For a general treatment, they proposed a full confidence interval
construction following Neyman \cite{JNeyman37}, exploiting the freedom in the ordering of values added to the acceptance regions. There are also approximate methods studied by others, see e.g. \cite{Cowan11}\cite{Cousins18}, to estimate confidence intervals and significance for experiments with small counts. 

There is another class of experiments in which the signal might produce a deficit with respect
to a known or measurable background. For experiments with small background count rate, observation of a signal deficit in the data is challenging. Apart from the well-known example of neutrino-oscillation experiments (see for example \cite{T2K2011}), another such case is the measurement of the hyperfine transition frequency of the ground-state antihydrogen atom \cite{Kuroda_NatComm2014}, which uses a Rabi-type in-flight magnetic spectroscopy method \cite{Diermaier17} to observe a double-dip resonance, and which suffers from the scarcity of antiatoms. Antimatter-based experiments have only recently started to reach a level where the first measurements of the properties of antihydrogen slowly became possible \cite{Alpha_NatPhys2011}\cite{Gabrielse_PRL2012}\cite{Alpha_1S2S}\cite{Alpha_Hyp}\cite{Alpha_LyA}, although still without sufficient sensitivity to perform precision CPT-invariance tests. The experimental difficulty for an antihydrogen in-flight experiment lies in producing enough, low temperature, spin-polarized antihydrogen atoms via atomic recombination and collisional processes such that they decay to ground-state during flight before reaching the various spectroscopic instruments and detectors. The recent collisional-radiative models predict the dominant background detected being due to highly-excited Rydberg-state antiatoms \cite{Radics14}, with the background rate itself being also very low or at least comparable to the signal. There exists, however, an almost full modeling of the antiatom in-flight transitions and magnetic forces on atomic momenta in a Rabi-like spectroscopic setup \cite{Lundmark15}, which may be used to construct confidence intervals for a number of model parameters (e.g. the transition frequency), possibly even in case of a low number of antiatoms. The aim of this work is to demonstrate that in the case of very low number of detected antihydrogen atoms, for the case of a signal deficit coupled with a known background level, it is possible to construct confidence intervals on the model parameters using the power of full statistical model inversion.  Traditional, approximate confidence interval construction methods offer reduced computational cost but might loose validity in this scenario. The method outlined in this paper uses the full Neyman-construction for the model parameter estimation in such a way that a Monte Carlo based simulation of the antihydrogen production and beam can be embedded in a straightforward way.

In order to quantitatively study the possibilities in low count experiments with small background and signal deficit, we first apply the full Neyman construction with the Feldman-Cousins likelihood ratio ordering approach to construct confidence intervals for a Poisson process with a known background and signal deficit. Then the multidimensional, Monte Carlo-based generalized version of the method is applied, taking the antihydrogen ground-state hyperfine measurement example, for an illustrative double-dip line shape. In this way, the construction of confidence regions for the peak position, width and strength of the resonance for various low data count scenarios is demonstrated. We find that within the assumptions of the expected antihydrogen velocity distribution and level population model, and assuming a line shape as that obtained from the hydrogen beam measurements, confidence intervals could be obtained for the transition frequency and other model parameters even from low statistics frequency scan data. Finally, we briefly outline also the general approach of confidence interval construction for a full simulation model.

%------------------------------------------------

\section{\label{sec:SingleCounting}Poisson process with known mean background and unknown signal deficit}

\begin{figure}[ht]
\includegraphics[width=0.7\linewidth]{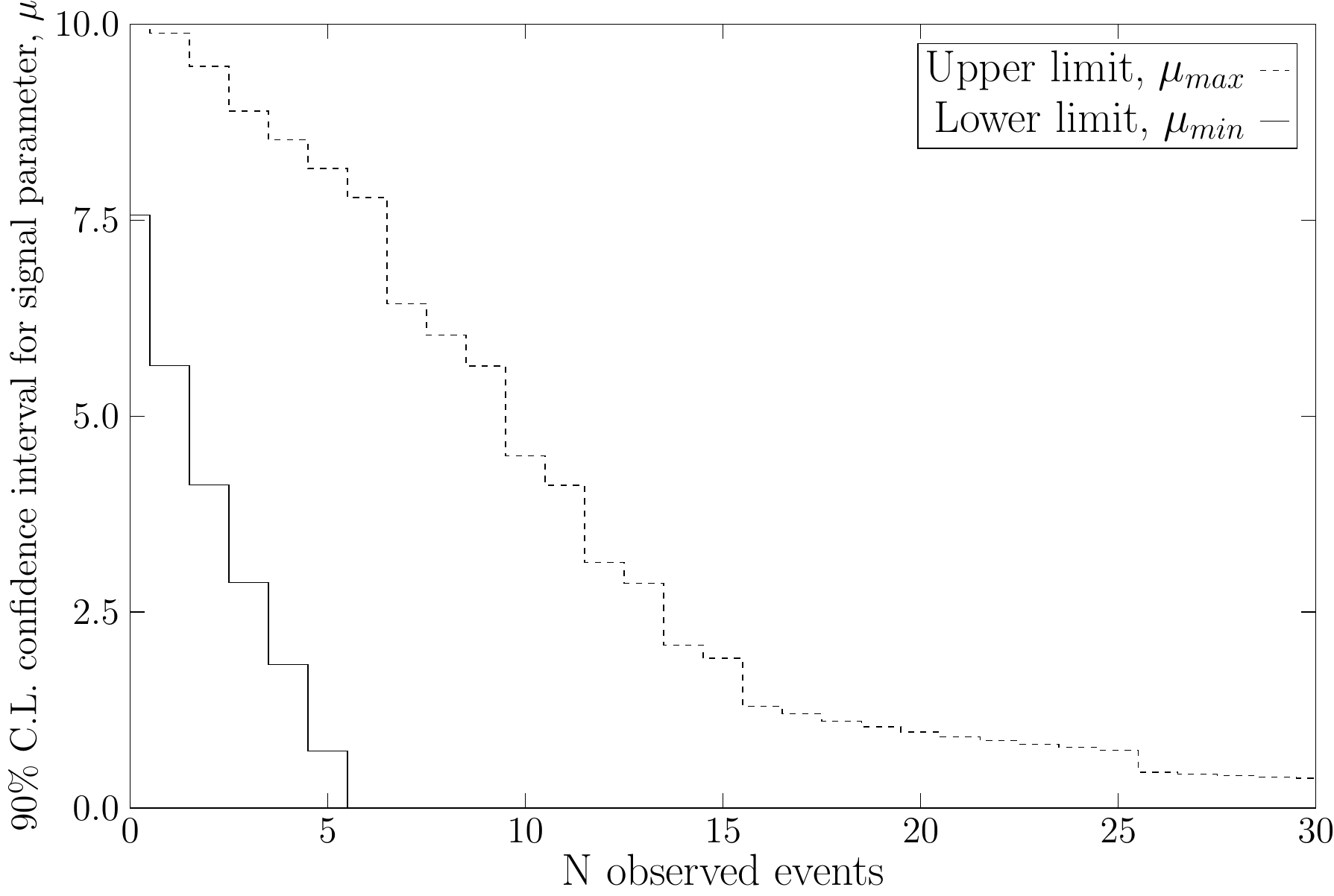}
\caption{Confidence belt for $90 \%$ C.L. confidence intervals for the unknown Poisson mean signal deficit parameter, $\mu$, in the presence background with known mean $b=10$ events.}
\label{fig:PoissonConfInt}
\end{figure}

We start with a brief outline of the model of a Poisson process with a known background and an unknown signal deficit, 
\begin{equation}
P(N|\mu) = \frac{(b - \mu)^{N}}{N!}e^{-(b-\mu)}, (b-\mu) \geq 0
\label{eq:poisson}
\end{equation}
where $N$ is the total number of events observed, $b$ is the known mean background, $\mu$ is the unknown mean signal parameter which characterizes the strength of the signal process. A value of $\mu = 0$ would correspond to a background-only case. In Eq.\eqref{eq:poisson} it is indicated explicitly that in this model the total mean, $b-\mu$, is required to be non-negative, that is negative mean is not part of the model. We perform a full Neyman-construction for this model, that is for a fixed background, $b$, and for each value of the unknown, physically allowed signal parameter, $\mu$, we analytically calculate the discrete probability distribution and determine an acceptance interval, $ [N_{
\mathrm{min}}, N_{\mathrm{max}}]$, such that it fulfils the requirement, 
\begin{equation}
P(N \in [N_{\mathrm{min}}, N_{\mathrm{max}}]|\mu) = \alpha
\end{equation}
where $\alpha$ is the desired confidence level (C.L.). From this one can find the intervals which have the property,
\begin{equation}
P(\mu \in [\mu_{\mathrm{min}}, \mu_{\mathrm{max}}]) = \alpha
\end{equation}
The construction is illustrated in Figure~\ref{fig:PoissonConfInt} for a fixed, known mean background of 10 events and $\alpha = 0.9$. The acceptance region is an interval $[N_{\mathrm{min}}, N_{\mathrm{max}}]$ for each value of $\mu$, which is obtained from the Feldman-Cousins ordering principle, which uses the quantity, $R = P(N|\mu)/P(N|\mu_{\mathrm{best}})$, as the basis of the ordering of adding the various values of $N$ to the acceptance regions. Here $\mu_{\mathrm{best}}$ is the value of $\mu$ which maximizes $P$ for the given observed $N$. The $90\%$ C.L. confidence interval of the unknown mean signal deficit parameter, $\mu$, for a given $N$, is obtained by finding those acceptance intervals that contain $N$, and quoting the smallest and largest value of $\mu$ from all of these intervals (It is noted that due to the discreteness of the probability distribution the intersecting intervals are not always simply connected, but this does not impact the results.) In contrast to the case of a (positive) signal excess with respect to the background, the intervals become upper limits for a larger number of observed events, which is expected. For very few observed events the intervals shift towards the known mean background level, by construction. That is, the model does not allow to obtain unphysically large signal deficit parameters, which would lead to negative values for the mean of the Poisson. 

%\begin{figure}[ht]
%\includegraphics[width=\linewidth]{plotPoisson_90CL_bkg10_Acc.pdf}
%\caption{Acceptance intervals, $[n_1, n_2]$, as a function of the signal deficit parameter, $\mu$, obtained during the Neyman-construction for the $90 \%$ C.L. confidence intervals, using the likelihood ratio ordering principle, for the case of the presence background with known mean $b=10$ events.}
%\label{fig:PoissonConfBelt}
%\end{figure}

To verify that obtained confidence intervals have the required property of $P(\mu \in [\mu_{\mathrm{min}}, \mu_{\mathrm{max}}]) = \alpha$, we perform Monte Carlo simulations under the assumption of a signal hypothesis and calculate in what fraction of the samples the confidence interval contains the true signal deficit parameter value. We find that the confidence intervals slightly overcover the parameter, which may only lead to conservatism in the interval estimates. 

For a general binned, multidimensional dataset, with $L$ bins, $\vec{N}=(N_{1}, ..., N_{L})$, and with $k$ parameters, $\vec{\mu} = (\mu_{1}, ..., \mu_{k})$ we use a slightly modified model, 
\begin{align}
& P(\vec{N}|\vec{\mu}) = \prod_{i}^{L} \frac{(f_{b,i} - f_{s,i}(\vec{\mu}))^{N_{i}}}{N_{i}!}e^{-(f_{b,i} - f_{s,i}(\vec{\mu}))}, \\
& f_{b,i} - f_{s,i}(\vec{\mu}) \geq 0
\label{eq:multibinpoisson}
\end{align}
where the background and signal related line shapes, $f_{b,i}$ and $f_{s,i}$, respectively, are included, and the parameters $\vec{\mu}$  illustrate a possibly multidimensional model with k parameters. An extended statistical model might also account for additional systematical uncertainties that potentially affect the shape. Again, we require that the mean of the individual Poisson processes must be non-negative in each bin. During construction, we use the likelihood ratio ordering principle,
\begin{equation}
R = P(\vec{N}|\vec{\mu})/P(\vec{N}|\vec{\mu}_{\mathrm{best}})
\end{equation}
to add values to the acceptance region for a given $\vec{\mu}$, in decreasing order of $R$.

%------------------------------------------------
\section{\label{sec:MultibinCounting}Application to the antihydrogen ground-state hyperfine transition resonant lineshape}
\subsection{\label{sec:Experiment}The experimental line shape with signal deficit}
The principle for the measurement of the ground-state hyperfine transition frequency of the antihydrogen atom is based on the Rabi-type magnetic resonance spectroscopy \cite{Rabi1938}. Upon production of a spin-polarized antihydrogen beam, a hyperfine transition in the ground-state antiatoms is driven by a cavity. The produced low-field-seeker and high-field-seeker quantum states are then focused and defocused, respectively, by a sextupole magnetic field. Due to the different populations of the quantum states an ensemble of antiatoms is subjected to various amounts of focusing and defocusing, which creates background and deficit signals in the final detector. In addition, there is an unpolarized fraction of quantum-states which will not be affected by the magnetic field.  The excited-state antiatoms reaching the detector represent the main source of background. The actual double-dip resonance line shape, for the experiment studied in the paper \cite{Diermaier17}, follows from the oscillating magnetic field pattern along the beam axis. The difficulty lies in the initial production of the spin-polarized antihydrogen beam. At the time of writing this paper, to the best knowledge of the authors, there is no reliable model prediction for the degree of polarization of the produced antihydrogen atoms. There is, however, a collisional-radiative model calculation of the yield of the ground-state antihydrogen atoms relative to that of the excited levels (as a function of various hypothetical plasma parameters)\cite{Radics14}, and an atomic code for the simulation of the in-flight excitation and for the effect of forces exerted by the magnetic field gradients on the quantum state-dependent magnetic moments of the antiatoms \cite{Lundmark15}. The line shape has been recently precisely measured for a hydrogen beam with the same spectroscopic beamline \cite{Diermaier17}, which allows us to define a model with background and signal deficit, and to simulate toy experiments for the confidence interval construction.

\begin{figure}[ht]
\includegraphics[width=0.7\linewidth]{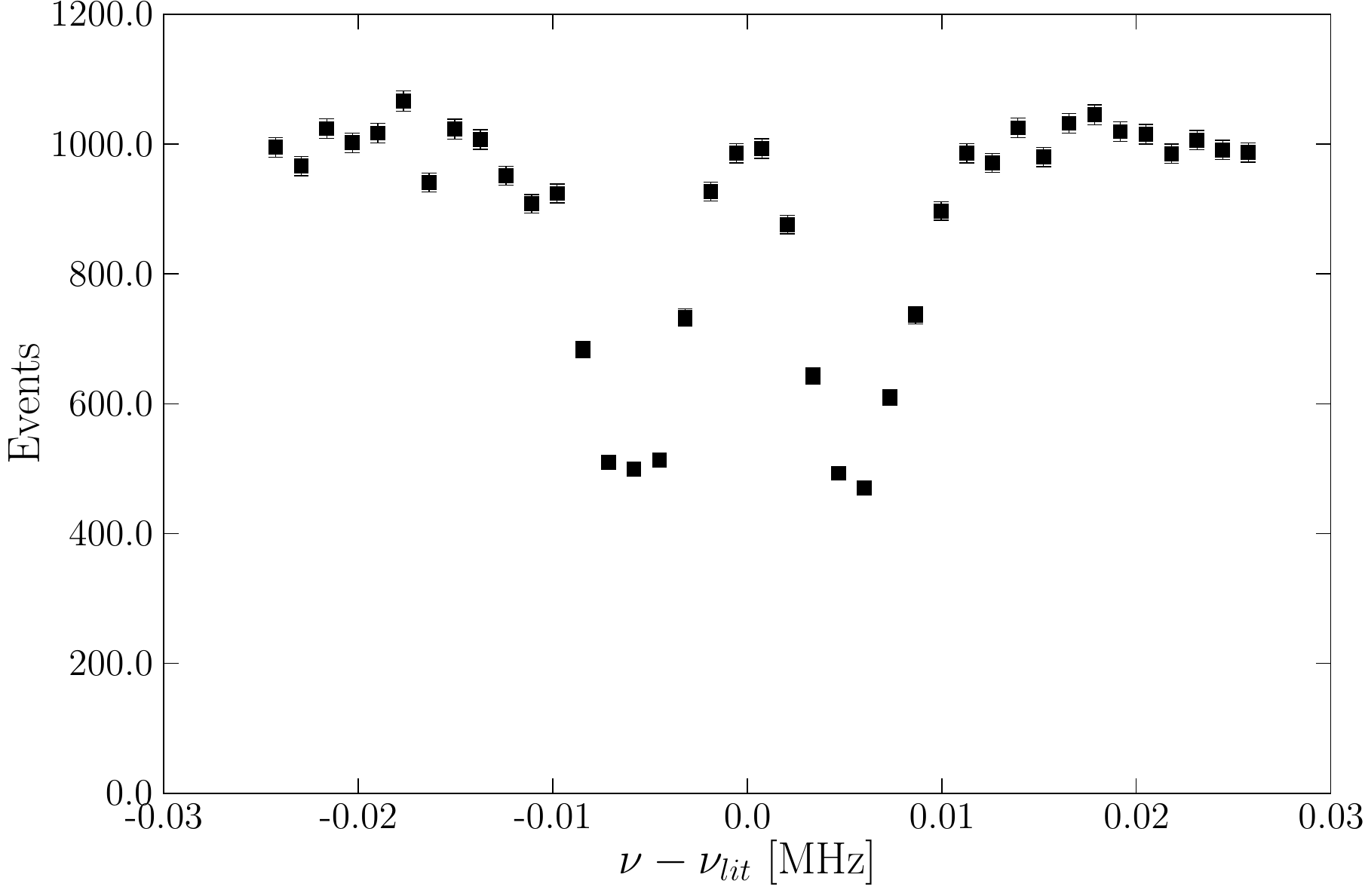}
\caption{Illustration of a double-dip resonance line shape, obtained from a simulated toy experiment with reasonably large statistics, on $b=1000$ mean background events per frequency scan point and using the nominal signal model outlined in the text.}
\label{fig:LineShapeBig}
\end{figure}

\begin{figure}[ht]
\includegraphics[width=0.7\linewidth]{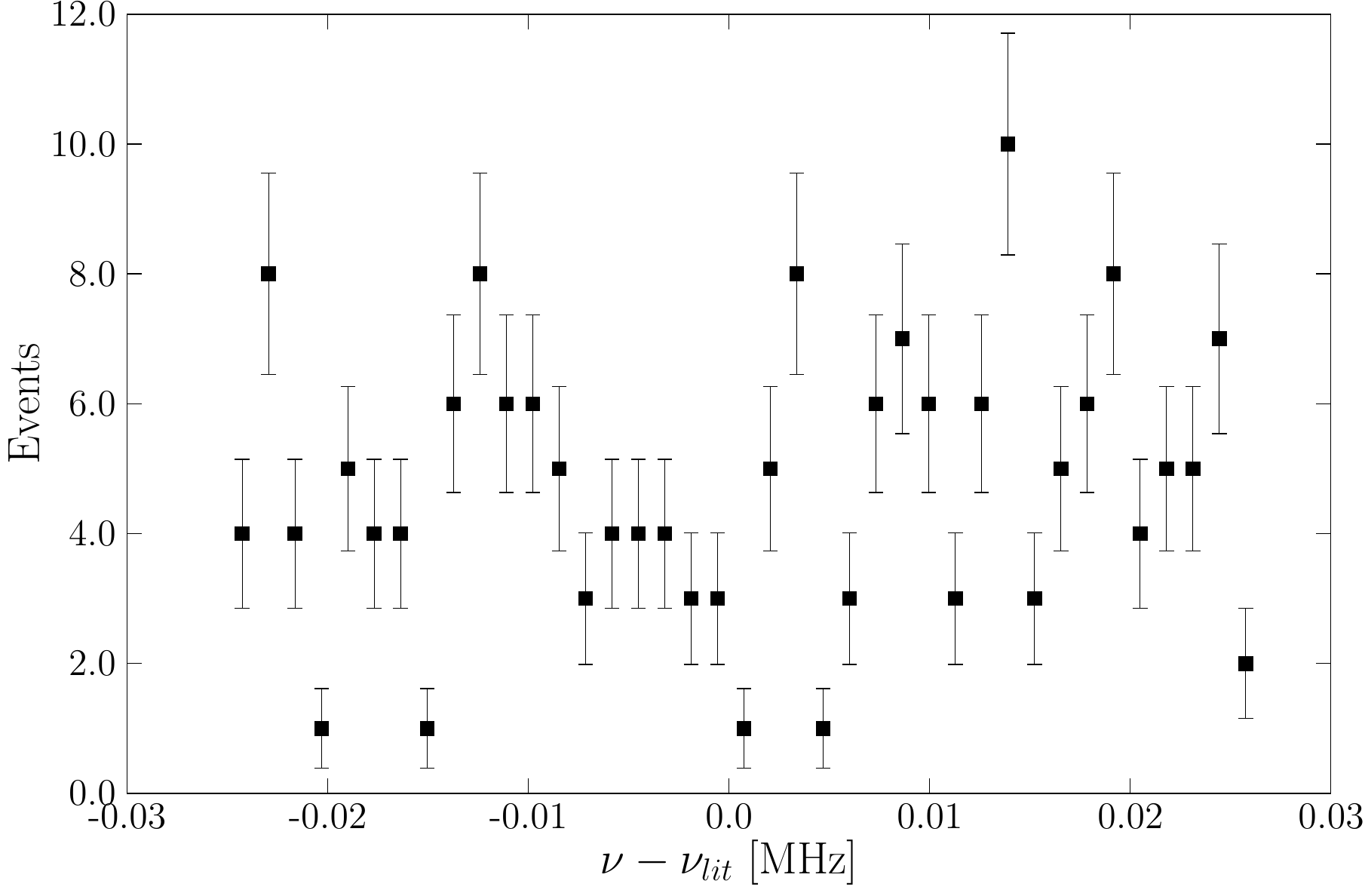}
\caption{Illustration of a double-dip resonance line shape, using the nominal model outlined in the text, obtained from a simulated toy experiment with low statistics, on average 5 background events per frequency scan point. }
\label{fig:LineShapeBkg5}
\end{figure}

The method is presented in the framework of the double-dip resonance line shape model, and afterwards, the possibilities for the extension towards a full, atomic code based calculation is discussed, see Section~\ref{sec:ExtendedModel}. The line shape simulation amounts to calculating the model prediction for the mean values, $f_{b,i}$ and $f_{s,i}$, for each frequency bin, $i$, and model parameter scan point, see Eq.\ref{eq:multibinpoisson}. As an illustration, a high statistics double-dip resonance line shape based toy dataset is shown in Fig.~\ref{fig:LineShapeBig}, assuming a presence of an average of 1000 background events per frequency scan point. The horizontal axis is defined as the amount of frequency detuning for the hyperfine transition with respect to the literature value of the transition frequency. For the purpose of this paper, we use a double-Gaussian to parametrize the double-dip resonance line shape, $f_{s}$, and a sinusoidal oscillating background, $f_{b}$. The amplitude of the Gaussian peaks is parametrized to be proportional relative to the background level, as we expect that an increase in the total antihydrogen production yield would manifest as a proportional increase in both the background and the signal amplitude. The two parameters of the line shape model are the transition frequency (in Figs.~\ref{fig:LineShapeBig} and ~\ref{fig:LineShapeBkg5} it was set to the literature value) and the common width of the Gaussian peaks, scanned in the analysis in a range of $\sigma \simeq 1-10$ kHz, which corresponds to the interaction time of the hydrogen atoms with the microwave fields, given their velocity, restricting the resonance line width. 

\begin{figure}[ht]
\includegraphics[width=0.7\linewidth]{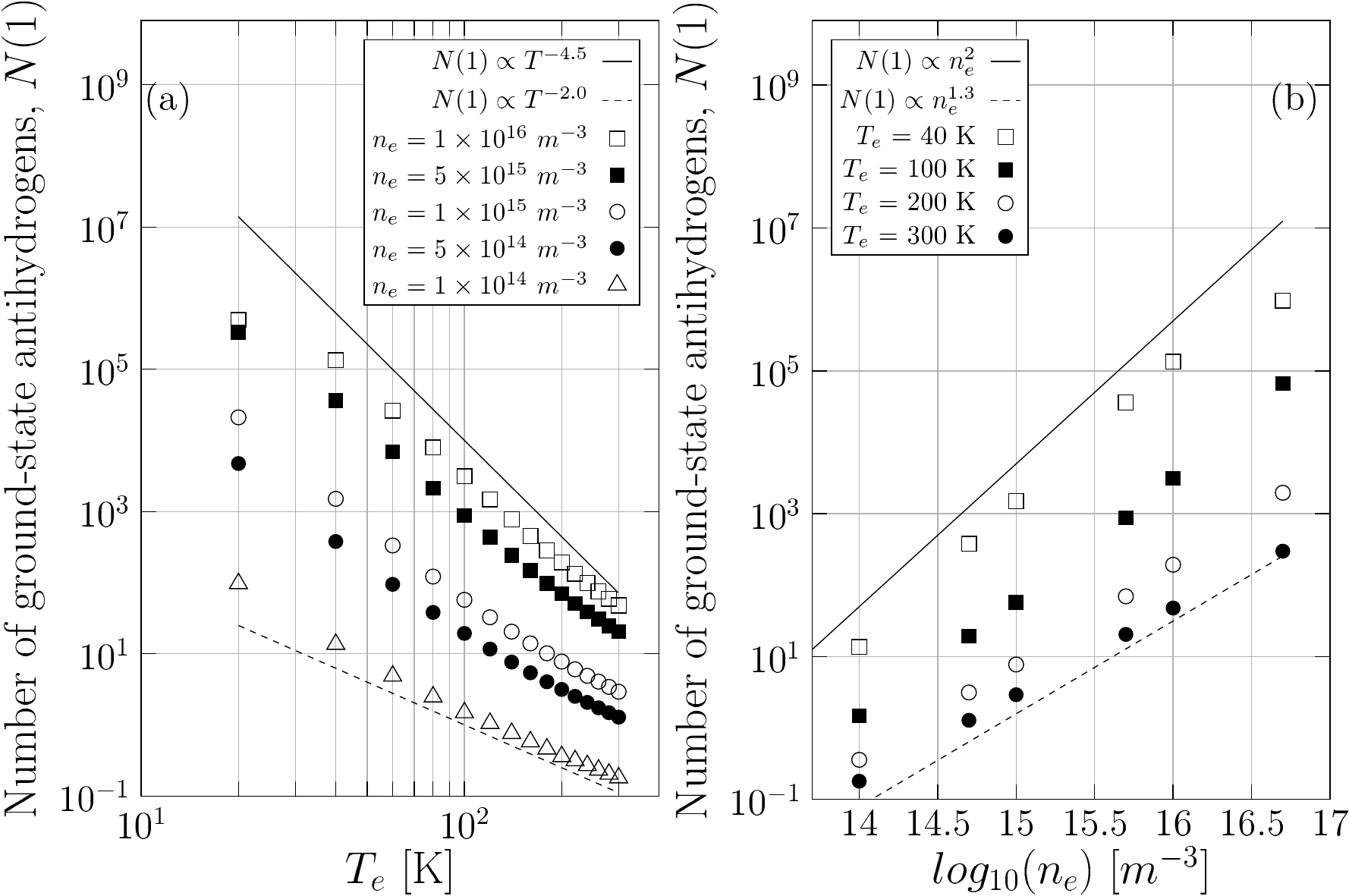}
\caption{Ground-state antihydrogen yield dependence on positron temperature (a) and density (b), after 10 $\mu s$ of mixing of a positron and antiproton plasma. Figure reproduced from \cite{Radics14}. }
\label{fig:maptempdens_flight}
\end{figure}

In practice, however, the count rate of antihydrogen atoms is rather low. An educated estimation of the total yield of ground-state antihydrogen atoms from a single passage of antiprotons through a positron plasma may be done based on the model developed in \cite{Radics14}. As illustrated in Fig.~\ref{fig:maptempdens_flight}, the model predictions suggest that the ground-state yield may change orders of magnitude depending on various plasma parameters, two of the most important ones being the positron plasma temperature and density. We take a conservative approach and assume a positron plasma temperature, $T_{e} \simeq 100$ K, and density, $n_{e} \simeq 10^{14} m^{-3}$. With this choice of the parameters, we arrive at a model prediction of 0.05 ground-state antihydrogen atoms per antiproton passage. Using the bouncing frequency of antiprotons inside a trapping potential of around an order of 100 kHz \cite{Enomoto10}, the approximate value of antiproton-positron plasma mixing time of around 30 seconds combined with a solid angle factor of $10^{-4}$  \cite{Kuroda_NatComm2014} , we arrive at a rough upper limit estimation of $N_{\mathrm{\bar{H}}} < 15$  ground-state antihydrogen atoms arriving at the detector during each antiproton-positron mixing experiment. Additional effects (such as detector efficiency, anisotropy of antihydrogen atom emission from the source, etc.) may significantly reduce this number, therefore a more realistic example of an experimental dataset is shown in Fig.~\ref{fig:LineShapeBkg5} assuming on average 5 background events per frequency point with the hypothesis of the presence of the signal. 
%Adding/fixing here
It is noted that the ground-state hyperfine transition signal might not be detectable due to various reasons: the antihydrogen atoms emerging from the source are not spin-polarized, their velocity is too high to decay to ground-state, etc. In the following, first we discuss the scenario with simulated signal present in the data. Afterwards, the scenario with the absence of detectable signal is also presented.

\subsection{\label{sec:ConfRegions}Construction of confidence regions}
\begin{figure}[ht]
\includegraphics[width=0.7\linewidth]{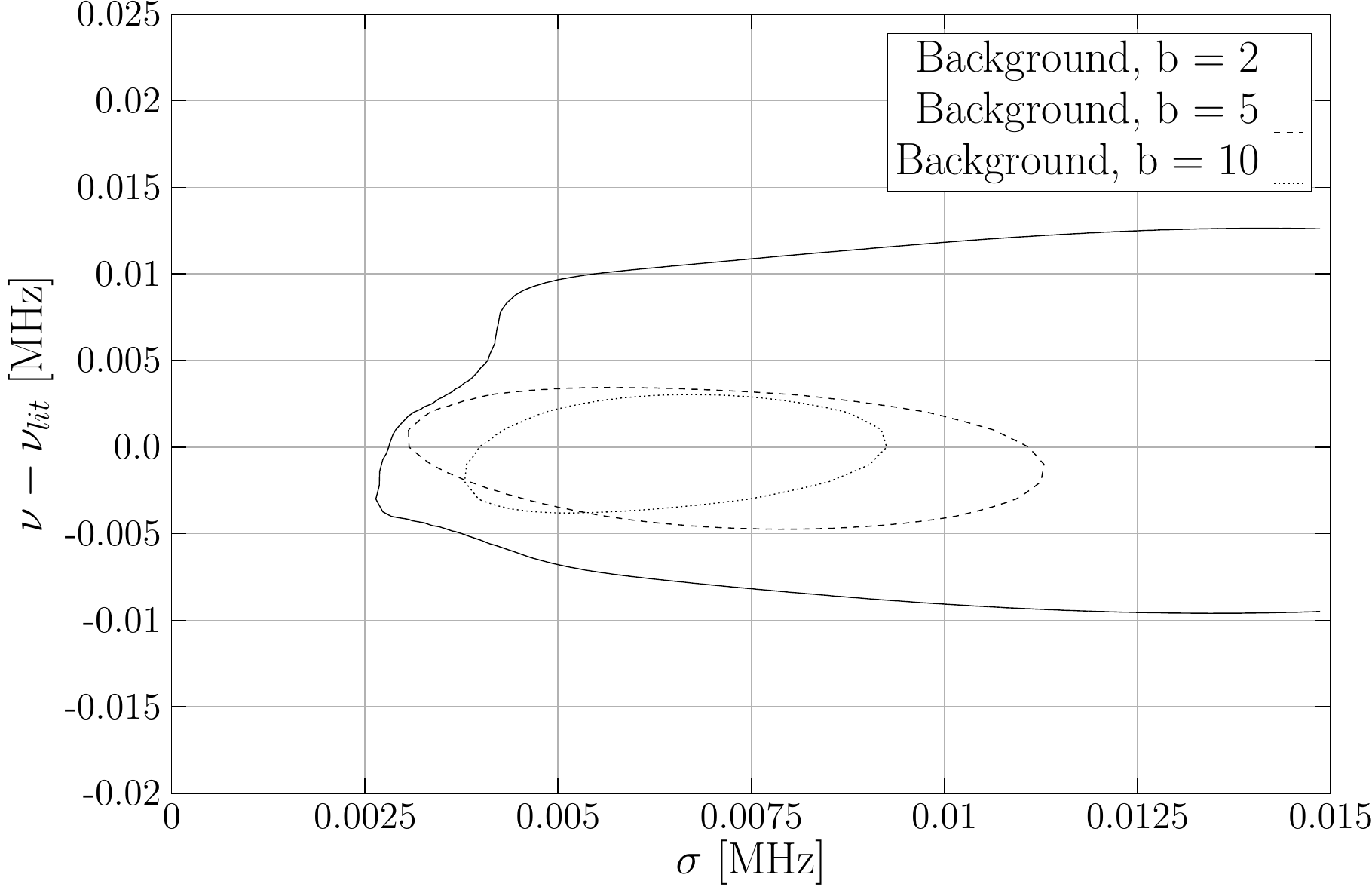}
\caption{Illustration of two-dimensional $90\%$ C.L.  confidence region boundaries for the hyperfine transition frequency and line width parameters, $\nu$ and $\sigma$, obtained for toy datasets generated with $\nu = \nu_{\mathrm{lit}}$ and $\sigma = 0.005$ MHz, in the presence of mean background $b = 2, 5$ and $10$ events per frequency points. }
\label{fig:ConfRegBkgs}
\end{figure}

\begin{figure}[ht]
\includegraphics[width=0.7\linewidth]{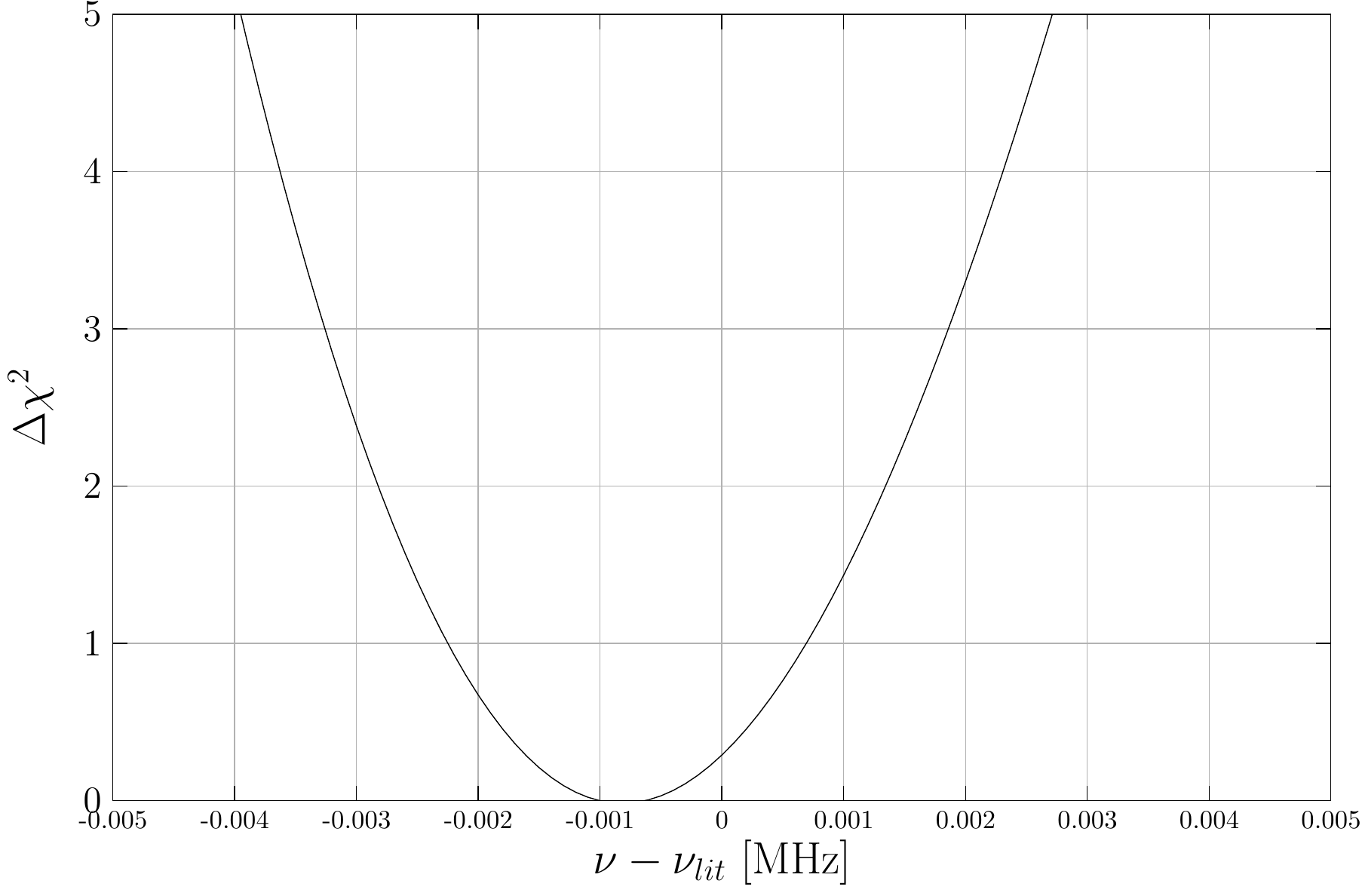}
\caption{Illustration of the $\Delta\chi^{2}$ profile around the minimum from the confidence intervals for the hyperfine transition frequency, for a toy dataset generated with $\nu = \nu_{\mathrm{lit}}$ and $\sigma = 0.005$ MHz, in the presence of mean background $b = 10$ events per frequency points. The vertical $\Delta\chi^{2}$ values correspond to the ones in Fig.~\ref{fig:ConfRegBkgs} for $b = 10$. }
\label{fig:Chi2ProfileBkg10}
\end{figure}

While the confidence intervals were obtained by a straightforward numerical calculation in Section~\ref{sec:SingleCounting}, for a multidimensional model and multi-bin data, the general approach is to use a Monte Carlo method. We briefly summarize here the solution of Feldman and Cousins \cite{FeldmanCousins98}. Instead of the total number of observed events, in this case, one has a number of possible $N$-sets (a frequency scan, such as  Fig.~\ref{fig:LineShapeBkg5}) for each parameter point in the two-dimensional $\nu-\sigma$ parameter space. In order to decide from the large possible $N$-sets which ones to allow in the acceptance region, we use the ratio $R = P(\vec{N}|\nu, \sigma)/P(\vec{N}|\nu_{\mathrm{best}}, \sigma_{\mathrm{best}} )$, or its equivalent version
\begin{equation}
\Delta\chi^{2} = -2 \cdot \ln \frac{P(\vec{N}|\nu, \sigma)}{P(\vec{N}|\nu_{\mathrm{best}}, \sigma_{\mathrm{best}} )},
\end{equation}
where $\nu$ and $\sigma$ are the parameter values in the two-dimensional space which are evaluated, and $\nu_{\mathrm{best}}$ and $\sigma_{\mathrm{best}}$ are the parameter values which give the highest probability, $P$, for the observed data set at that parameter point. One can then simulate a large number of toy experiments for each 2D parameter space point, and calculate $\Delta\chi^{2}$ for each of these experiments. From the large number of toy datasets, the $\Delta\chi^{2}$ values are used as the ordering quantity. At each parameter point a critical value, $\Delta\chi^{2}_{c}$, can be determined, such that $\Delta\chi^{2} < \Delta\chi^{2}_{c}$ for $\alpha$ fraction of the simulated experiments. Having the critical $\Delta\chi^{2}_c(\nu, \sigma)$ value map thus constructed, for any new dataset, $\Delta\chi^{2}_{d}$ and for each point in the parameter space, the confidence region can be determined by the condition that $\Delta\chi^{2}_{d}(\nu, \sigma) < \Delta\chi^{2}_{c}(\nu, \sigma)$.
The result of applying this method on a random low statistic toy dataset with mean background values of $b = 2, 5$ and $10$ and with the signal present is illustrated in Fig.~\ref{fig:ConfRegBkgs}. The profile of the $\Delta\chi^{2}$ values for $b=10$, obtained as the set of values at the best $\sigma$ width parameter (given the data)  is shown in Fig.~\ref{fig:Chi2ProfileBkg10}, as a function of the hyperfine transition frequency. The example illustrates the obtainable confidence regions from early data, assuming the signal model discussed. A confidence interval, such as the one show in  Fig.~\ref{fig:Chi2ProfileBkg10}, would correspond to an at least three orders of magnitude worse relative precision on the ground-state hyperfine transition frequency than that recently obtained in the hydrogen beam case with the same spectroscopic beam line. The method discussed in this paper however could allow optimizing the sensitivity of the antihydrogen experiment to find the best experimental conditions or to quantify the amount of data to be collected leading to a target precision. 

In case the signal has not been observed yet, the significance of the observed data can be also be quantified within the same statistical calculation by extending the parameter space with a signal strength parameter, $s$, that is $P \equiv P(\vec{N}|\nu, \sigma, s)$. The previously mentioned likelihood ratio (or $\Delta\chi^{2}$) then can be calculated for the background-only hypothesis and used as a test statistic to quantify the level agreement between the data and the null-hypothesis, following e.g. \cite{Cowan11}.

\begin{figure}[ht]
\includegraphics[width=0.7\linewidth]{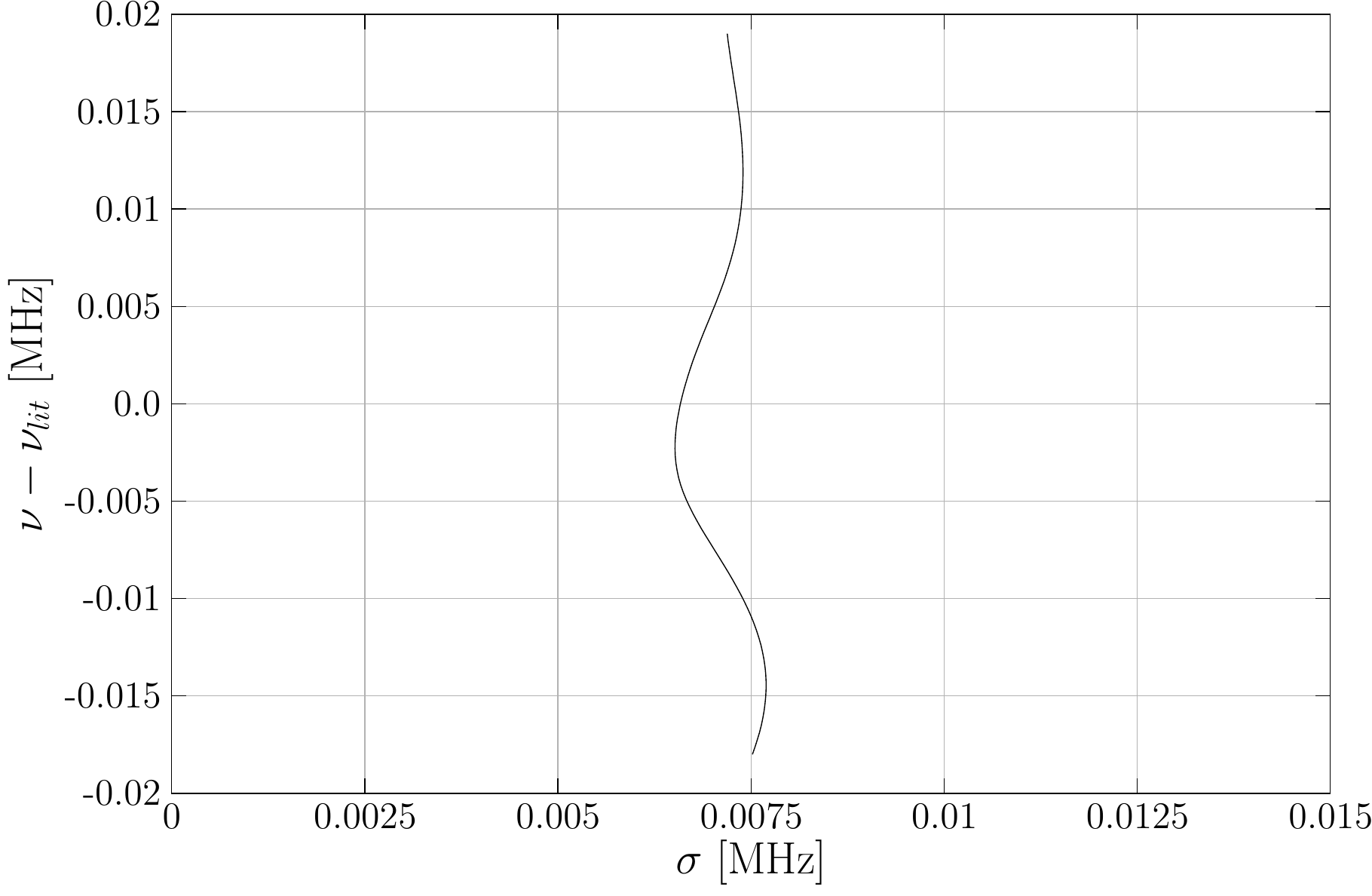}
\caption{Illustration of two-dimensional sensitivity boundary for the hyperfine transition frequency and line width parameters, $\nu$ and $\sigma$, obtained from an ensemble of toy datasets with mean background $b=10$ and zero signal. The confidence region is to the right of the boundary.}
\label{fig:Sensitivity}
\end{figure}

The sensitivity of this procedure is defined as the average upper limit that would be obtained from an ensemble of toy experiments at a given mean $b$ background level, but with no signal present. As shown in Fig.~\ref{fig:Sensitivity} we obtained a rather broad $90\%$ C.L. sensitivity curve for $b=10$ mean background level, for the two-dimensional model parameter space, $(\nu, \sigma)$. The confidence region is to the right of the curve. Due to the lack of signal, the average upper limit does not favor any particular transition frequency values, however, due to fluctuations in the data and due to the rather flat background, larger values of the Gaussian width parameters cannot be entirely excluded. 

%------------------------------------------------

The statistical coverage of the method over the two-dimensional parameter space has been also calculated by generating a large number of toy experiments and counting how many of the cases the true parameter is contained in the obtained confidence region at the desired $\alpha$. Similar to the case of a single Poisson process, we get slight overcover, $\sim 92\%$, when constructing critical $\Delta\chi^{2}_{c}$ for $\alpha = 0.9$. It is uniform over the whole parameter space, except towards the edge of the width parameter, $\sigma$. The extremely low or high $\sigma$ values belong to unphysical parameter regions and therefore the scanned parameter space does not include such values beyond the edges.

%------------------------------------------------

\section{\label{sec:ExtendedModel}Confidence intervals for extended models}
The calculation of confidence intervals can be extended to include the full modeling of the atomic processes and the effect of transitions and magnetic forces on the various quantum states of the antihydrogen atoms during their passage through the spectroscopic beamline. To obtain the initial antihydrogen level population distribution a set of coupled differential rate equations need to be solved, in which the rate coefficients are calculated from atomic scattering codes. Level population distributions can be produced for a set of various parameters and ranges and used as the initial point of the full simulation \cite{Radics14}. Although the precise non-neutral plasma conditions are difficult to measure, it may still be possible to account for them as phenomenological model parameters. In order to obtain the theoretical line shape for a particular beam, the state conversion probability needs to be calculated as a function of the driving frequency and the various fields \cite{Diermaier17}. Two of the interesting extensions to the model could be the addition of the ratio of the low-field-seeker to high-field-seeker quantum states and the velocity distribution of the antihydrogen atoms. Both impact the hyperfine resonance line shape \cite{DiermaierThesis}. While the confidence regions for the hyperfine transition frequency and line width have been demonstrated to cover the true parameters successfully, the sensitivity to the above mentioned additional parameters may not be so obvious. Nevertheless, confidence intervals can also be constructed for each parameter of these model scenarios in a straightforward way. For a particular parameter vector the full simulation would produce a model prediction for the amount of mean background and signal, $f_{b,i}$ and $f_{s,i}$ with $f_{b,i}-f_{s,i} \geq 0$ , in each frequency bin. The scan of the multidimensional parameter space can be performed in a fully parallel computation. The single value needed to characterize each model parameter point within the $k$-dimensional parameter space is the critical value, $\Delta\chi^{2}_{c}(\nu, \sigma, ...)$, for the $\alpha$ confidence level. A scan with the assumed known background level may reveal how much data is to be collected to obtain meaningful confidence regions from the data for each of these parameters. %In an analog way, one would be able to construct a critical $\Delta\chi^{2}_{\mathrm{c}}(\nu, \mu, s)$ map by simulating a large number of toy experiments around model prediction for the mean values, $f_{b,i}(\nu,\mu, s)$ and $f_{s,i}(\nu, \mu, s)$, using a full calculation.

%As an example, we show confidence intervals for an extended model, in which the amplitude of the double-dip resonance is also allowed to change with respect to the background level. The construction of the confidence region was performed in the three-dimensional parameter space, $(\nu, \sigma, s)$, where $s$ is a factor characterizing the magnitude of the resonance amplitude. This would correspond to a scenario when the amount of low-field-seeker states are allowed to vary with respect to a nominal fraction. Because this case is three-dimensional we show only the two-dimensional projections into the subspaces separately for a toy dataset generated with half the signal strength of the nominal signal model.

%------------------------------------------------

\section{\label{sec:Conclusion}Conclusion}
In this paper, we presented the Feldman-Cousins confidence interval construction applied to the case of a signal deficit, in particular to the antihydrogen hyperfine transition line shape. Given the low expected experimental rate of the ground-state antihydrogen atoms, and due to the complicated mechanisms of their production, there might be a lot of uncertainty associated with the actual line shape. We showed an example of confidence interval construction for two antihydrogen model parameters, the hyperfine transition frequency and the line width, both related either to the antihydrogen atom or to the beam properties. The discussed approach potentially allows extraction of a number of additional atomic and beam related quantities of interest, by performing parameter scans via a full atomic code and simulations and constructing intervals using the Monte Carlo method. In contrast to the fitting procedure with a numerical formula, used in the case of the high count rate hydrogen beam data, the method discussed in this paper uses a full Neyman-construction to obtain confidence intervals of model parameters. This allows the embedding of a full simulation of the beam line and some of the antihydrogen properties to be estimated from very early, low statistics data. 

%------------------------------------------------
\section*{Data accessibility statement}
This work does not have any experimental data.

\section*{Competing interests statement}
We have no competing interests.

\section*{Authors' contributions}
BR constructed the statistical model, performed the simulation, interpreted the computational results, and wrote the paper.

\section*{Funding statement}
This work was supported by ETH Z\"urich.

%------------------------------------------------
\begin{acknowledgments}

We gratefully acknowledge conversations with D. A. Cooke and D. Sgalaberna.  \\

\end{acknowledgments}
%------------------------------------------------

\bibliography{Bibliography}

\begin{thebibliography}{17}%
\makeatletter
\providecommand \@ifxundefined [1]{%
 \@ifx{#1\undefined}
}%
\providecommand \@ifnum [1]{%
 \ifnum #1\expandafter \@firstoftwo
 \else \expandafter \@secondoftwo
 \fi
}%
\providecommand \@ifx [1]{%
 \ifx #1\expandafter \@firstoftwo
 \else \expandafter \@secondoftwo
 \fi
}%
\providecommand \natexlab [1]{#1}%
\providecommand \enquote  [1]{``#1''}%
\providecommand \bibnamefont  [1]{#1}%
\providecommand \bibfnamefont [1]{#1}%
\providecommand \citenamefont [1]{#1}%
\providecommand \href@noop [0]{\@secondoftwo}%
\providecommand \href [0]{\begingroup \@sanitize@url \@href}%
\providecommand \@href[1]{\@@startlink{#1}\@@href}%
\providecommand \@@href[1]{\endgroup#1\@@endlink}%
\providecommand \@sanitize@url [0]{\catcode `\\12\catcode `\$12\catcode
  `\&12\catcode `\#12\catcode `\^12\catcode `\_12\catcode `\%12\relax}%
\providecommand \@@startlink[1]{}%
\providecommand \@@endlink[0]{}%
\providecommand \url  [0]{\begingroup\@sanitize@url \@url }%
\providecommand \@url [1]{\endgroup\@href {#1}{\urlprefix }}%
\providecommand \urlprefix  [0]{URL }%
\providecommand \Eprint [0]{\href }%
\providecommand \doibase [0]{http://dx.doi.org/}%
\providecommand \selectlanguage [0]{\@gobble}%
\providecommand \bibinfo  [0]{\@secondoftwo}%
\providecommand \bibfield  [0]{\@secondoftwo}%
\providecommand \translation [1]{[#1]}%
\providecommand \BibitemOpen [0]{}%
\providecommand \bibitemStop [0]{}%
\providecommand \bibitemNoStop [0]{.\EOS\space}%
\providecommand \EOS [0]{\spacefactor3000\relax}%
\providecommand \BibitemShut  [1]{\csname bibitem#1\endcsname}%
\let\auto@bib@innerbib\@empty
%</preamble>
\bibitem [{\citenamefont {Feldman}\ and\ \citenamefont
  {Cousins}(1998)}]{FeldmanCousins98}%
  \BibitemOpen
  \bibfield  {author} {\bibinfo {author} {\bibfnamefont {G.~J.}\ \bibnamefont
  {Feldman}}\ and\ \bibinfo {author} {\bibfnamefont {R.~D.}\ \bibnamefont
  {Cousins}},\ }\href@noop {} {\bibfield  {journal} {\bibinfo  {journal} {Phys.
  Rev. D, 3873}\ }\textbf {\bibinfo {volume} {57}},\ \bibinfo {pages} {3873}
  (\bibinfo {year} {1998})}\BibitemShut {NoStop}%
\bibitem [{\citenamefont {Neyman}(1937)}]{JNeyman37}%
  \BibitemOpen
  \bibfield  {author} {\bibinfo {author} {\bibfnamefont {J.}~\bibnamefont
  {Neyman}},\ }\href {\doibase 10.1098/rsta.1937.0005} {\bibfield  {journal}
  {\bibinfo  {journal} {Philos. Trans. R. Soc. London}\ }\textbf {\bibinfo
  {volume} {A236}},\ \bibinfo {pages} {333} (\bibinfo {year}
  {1937})}\BibitemShut {NoStop}%
\bibitem [{\citenamefont {Cowan}\ \emph {et~al.}(2011)\citenamefont {Cowan},
  \citenamefont {Cranmer}, \citenamefont {Gross},\ and\ \citenamefont
  {Vitells}}]{Cowan11}%
  \BibitemOpen
  \bibfield  {author} {\bibinfo {author} {\bibfnamefont {G.}~\bibnamefont
  {Cowan}}, \bibinfo {author} {\bibfnamefont {K.}~\bibnamefont {Cranmer}},
  \bibinfo {author} {\bibfnamefont {E.}~\bibnamefont {Gross}}, \ and\ \bibinfo
  {author} {\bibfnamefont {O.}~\bibnamefont {Vitells}},\ }\href {\doibase
  10.1140/epjc/s10052-011-1554-0} {\bibfield  {journal} {\bibinfo  {journal}
  {Eur. Phys. J. C}\ }\textbf {\bibinfo {volume} {71}},\ \bibinfo {pages}
  {1554} (\bibinfo {year} {2011})}\BibitemShut {NoStop}%
\bibitem [{\citenamefont {Cousins}\ \emph {et~al.}(2008)\citenamefont
  {Cousins}, \citenamefont {Linnemann},\ and\ \citenamefont
  {Tucker}}]{Cousins18}%
  \BibitemOpen
  \bibfield  {author} {\bibinfo {author} {\bibfnamefont {R.~D.}\ \bibnamefont
  {Cousins}}, \bibinfo {author} {\bibfnamefont {J.~T.}\ \bibnamefont
  {Linnemann}}, \ and\ \bibinfo {author} {\bibfnamefont {J.}~\bibnamefont
  {Tucker}},\ }\href@noop {} {\bibfield  {journal} {\bibinfo  {journal} {Nucl.
  Instr. Meth. A}\ }\textbf {\bibinfo {volume} {595}},\ \bibinfo {pages} {480}
  (\bibinfo {year} {2008})}\BibitemShut {NoStop}%
\bibitem [{\citenamefont {Abe}\ \emph {et~al.}(2011)\citenamefont {Abe} \emph
  {et~al.}}]{T2K2011}%
  \BibitemOpen
  \bibfield  {author} {\bibinfo {author} {\bibfnamefont {K.}~\bibnamefont
  {Abe}} \emph {et~al.},\ }\href@noop {} {\bibfield  {journal} {\bibinfo
  {journal} {Phys. Rev. Lett.}\ }\textbf {\bibinfo {volume} {107}},\ \bibinfo
  {pages} {041801} (\bibinfo {year} {2011})}\BibitemShut {NoStop}%
\bibitem [{\citenamefont {Kuroda}\ \emph {et~al.}(2014)\citenamefont {Kuroda}
  \emph {et~al.}}]{Kuroda_NatComm2014}%
  \BibitemOpen
  \bibfield  {author} {\bibinfo {author} {\bibfnamefont {N.}~\bibnamefont
  {Kuroda}} \emph {et~al.},\ }\href@noop {} {\bibfield  {journal} {\bibinfo
  {journal} {Nature Communications}\ }\textbf {\bibinfo {volume} {5}},\
  \bibinfo {pages} {3089} (\bibinfo {year} {2014})}\BibitemShut {NoStop}%
\bibitem [{\citenamefont {Diermaier}\ \emph {et~al.}(2017)\citenamefont
  {Diermaier} \emph {et~al.}}]{Diermaier17}%
  \BibitemOpen
  \bibfield  {author} {\bibinfo {author} {\bibfnamefont {M.}~\bibnamefont
  {Diermaier}} \emph {et~al.},\ }\href {\doibase 10.1038/ncomms15749}
  {\bibfield  {journal} {\bibinfo  {journal} {Nature Communications}\ }\textbf
  {\bibinfo {volume} {8}},\ \bibinfo {pages} {15749} (\bibinfo {year}
  {2017})}\BibitemShut {NoStop}%
\bibitem [{\citenamefont {Andresen}\ \emph {et~al.}(2011)\citenamefont
  {Andresen} \emph {et~al.}}]{Alpha_NatPhys2011}%
  \BibitemOpen
  \bibfield  {author} {\bibinfo {author} {\bibfnamefont {G.~B.}\ \bibnamefont
  {Andresen}} \emph {et~al.},\ }\href@noop {} {\bibfield  {journal} {\bibinfo
  {journal} {Nat. Phys.}\ }\textbf {\bibinfo {volume} {7}},\ \bibinfo {pages}
  {558} (\bibinfo {year} {2011})}\BibitemShut {NoStop}%
\bibitem [{\citenamefont {Gabrielse}\ \emph {et~al.}(2012)\citenamefont
  {Gabrielse}, \citenamefont {Kalra}, \citenamefont {Kolthammer}, \citenamefont
  {McConnell}, \citenamefont {Richerme}, \citenamefont {Grzonka}, \citenamefont
  {Oelert}, \citenamefont {Sefzick}, \citenamefont {Zielinski}, \citenamefont
  {Fitzakerley}, \citenamefont {George}, \citenamefont {Hessels}, \citenamefont
  {Storry}, \citenamefont {Weel}, \citenamefont {Mullers},\ and\ \citenamefont
  {Walz}}]{Gabrielse_PRL2012}%
  \BibitemOpen
  \bibfield  {author} {\bibinfo {author} {\bibfnamefont {G.}~\bibnamefont
  {Gabrielse}}, \bibinfo {author} {\bibfnamefont {R.}~\bibnamefont {Kalra}},
  \bibinfo {author} {\bibfnamefont {W.}~\bibnamefont {Kolthammer}}, \bibinfo
  {author} {\bibfnamefont {R.}~\bibnamefont {McConnell}}, \bibinfo {author}
  {\bibfnamefont {P.}~\bibnamefont {Richerme}}, \bibinfo {author}
  {\bibfnamefont {D.}~\bibnamefont {Grzonka}}, \bibinfo {author} {\bibfnamefont
  {W.}~\bibnamefont {Oelert}}, \bibinfo {author} {\bibfnamefont
  {T.}~\bibnamefont {Sefzick}}, \bibinfo {author} {\bibfnamefont
  {M.}~\bibnamefont {Zielinski}}, \bibinfo {author} {\bibfnamefont
  {D.}~\bibnamefont {Fitzakerley}}, \bibinfo {author} {\bibfnamefont
  {M.}~\bibnamefont {George}}, \bibinfo {author} {\bibfnamefont
  {E.}~\bibnamefont {Hessels}}, \bibinfo {author} {\bibfnamefont
  {C.}~\bibnamefont {Storry}}, \bibinfo {author} {\bibfnamefont
  {M.}~\bibnamefont {Weel}}, \bibinfo {author} {\bibfnamefont {A.}~\bibnamefont
  {Mullers}}, \ and\ \bibinfo {author} {\bibfnamefont {J.}~\bibnamefont
  {Walz}},\ }\href@noop {} {\bibfield  {journal} {\bibinfo  {journal} {Phys.
  Rev. Lett.}\ }\textbf {\bibinfo {volume} {108}},\ \bibinfo {pages} {113002}
  (\bibinfo {year} {2012})}\BibitemShut {NoStop}%
\bibitem [{\citenamefont {Ahmadi}\ \emph
  {et~al.}(2017{\natexlab{a}})\citenamefont {Ahmadi} \emph
  {et~al.}}]{Alpha_1S2S}%
  \BibitemOpen
  \bibfield  {author} {\bibinfo {author} {\bibfnamefont {M.}~\bibnamefont
  {Ahmadi}} \emph {et~al.},\ }\href@noop {} {\bibfield  {journal} {\bibinfo
  {journal} {Nature}\ }\textbf {\bibinfo {volume} {541}},\ \bibinfo {pages}
  {506} (\bibinfo {year} {2017}{\natexlab{a}})}\BibitemShut {NoStop}%
\bibitem [{\citenamefont {Ahmadi}\ \emph
  {et~al.}(2017{\natexlab{b}})\citenamefont {Ahmadi} \emph
  {et~al.}}]{Alpha_Hyp}%
  \BibitemOpen
  \bibfield  {author} {\bibinfo {author} {\bibfnamefont {M.}~\bibnamefont
  {Ahmadi}} \emph {et~al.},\ }\href@noop {} {\bibfield  {journal} {\bibinfo
  {journal} {Nature}\ }\textbf {\bibinfo {volume} {548}},\ \bibinfo {pages}
  {66} (\bibinfo {year} {2017}{\natexlab{b}})}\BibitemShut {NoStop}%
\bibitem [{\citenamefont {Ahmadi}\ \emph {et~al.}(2018)\citenamefont {Ahmadi}
  \emph {et~al.}}]{Alpha_LyA}%
  \BibitemOpen
  \bibfield  {author} {\bibinfo {author} {\bibfnamefont {M.}~\bibnamefont
  {Ahmadi}} \emph {et~al.},\ }\href {\doibase 10.1038/s41586-018-0435-1}
  {\bibfield  {journal} {\bibinfo  {journal} {Nature}\ } (\bibinfo {year}
  {2018}),\ 10.1038/s41586-018-0435-1}\BibitemShut {NoStop}%
\bibitem [{\citenamefont {Radics}\ \emph {et~al.}(2014)\citenamefont {Radics},
  \citenamefont {Murtagh}, \citenamefont {Yamazaki},\ and\ \citenamefont
  {Robicheaux}}]{Radics14}%
  \BibitemOpen
  \bibfield  {author} {\bibinfo {author} {\bibfnamefont {B.}~\bibnamefont
  {Radics}}, \bibinfo {author} {\bibfnamefont {D.~J.}\ \bibnamefont {Murtagh}},
  \bibinfo {author} {\bibfnamefont {Y.}~\bibnamefont {Yamazaki}}, \ and\
  \bibinfo {author} {\bibfnamefont {F.}~\bibnamefont {Robicheaux}},\
  }\href@noop {} {\bibfield  {journal} {\bibinfo  {journal} {Phys. Rev. A}\
  }\textbf {\bibinfo {volume} {90}},\ \bibinfo {pages} {032704} (\bibinfo
  {year} {2014})}\BibitemShut {NoStop}%
\bibitem [{\citenamefont {Lundmark}\ \emph {et~al.}(2015)\citenamefont
  {Lundmark}, \citenamefont {Malbrunot}, \citenamefont {Nagata}, \citenamefont
  {Radics}, \citenamefont {Sauerzopf},\ and\ \citenamefont
  {Widmann}}]{Lundmark15}%
  \BibitemOpen
  \bibfield  {author} {\bibinfo {author} {\bibfnamefont {R.}~\bibnamefont
  {Lundmark}}, \bibinfo {author} {\bibfnamefont {C.}~\bibnamefont {Malbrunot}},
  \bibinfo {author} {\bibfnamefont {Y.}~\bibnamefont {Nagata}}, \bibinfo
  {author} {\bibfnamefont {B.}~\bibnamefont {Radics}}, \bibinfo {author}
  {\bibfnamefont {C.}~\bibnamefont {Sauerzopf}}, \ and\ \bibinfo {author}
  {\bibfnamefont {E.}~\bibnamefont {Widmann}},\ }\href@noop {} {\bibfield
  {journal} {\bibinfo  {journal} {J. of Phys. B: At. Mol. Opt. Phys.}\ }\textbf
  {\bibinfo {volume} {48}},\ \bibinfo {pages} {184001} (\bibinfo {year}
  {2015})}\BibitemShut {NoStop}%
\bibitem [{\citenamefont {Rabi}\ \emph {et~al.}(1938)\citenamefont {Rabi},
  \citenamefont {Zacharias}, \citenamefont {Millman},\ and\ \citenamefont
  {Kusch}}]{Rabi1938}%
  \BibitemOpen
  \bibfield  {author} {\bibinfo {author} {\bibfnamefont {I.~I.}\ \bibnamefont
  {Rabi}}, \bibinfo {author} {\bibfnamefont {J.~R.}\ \bibnamefont {Zacharias}},
  \bibinfo {author} {\bibfnamefont {S.}~\bibnamefont {Millman}}, \ and\
  \bibinfo {author} {\bibfnamefont {P.}~\bibnamefont {Kusch}},\ }\href@noop {}
  {\bibfield  {journal} {\bibinfo  {journal} {Phys. Rev.}\ }\textbf {\bibinfo
  {volume} {53}},\ \bibinfo {pages} {318} (\bibinfo {year} {1938})}\BibitemShut
  {NoStop}%
\bibitem [{\citenamefont {Enomoto}\ \emph {et~al.}(2010)\citenamefont
  {Enomoto}, \citenamefont {Kuroda}, \citenamefont {Michishio}, \citenamefont
  {Kim}, \citenamefont {Higaki}, \citenamefont {Nagata}, \citenamefont {Kanai},
  \citenamefont {Torii}, \citenamefont {Corradini}, \citenamefont {Leali},
  \citenamefont {Lodi-Rizzini}, \citenamefont {Mascagna}, \citenamefont
  {Venturelli}, \citenamefont {Zurlo}, \citenamefont {Fujii}, \citenamefont
  {Ohtsuka}, \citenamefont {Tanaka}, \citenamefont {Imao}, \citenamefont
  {Nagashima}, \citenamefont {Matsuda}, \citenamefont {Juhasz}, \citenamefont
  {Mohri},\ and\ \citenamefont {Yamazaki}}]{Enomoto10}%
  \BibitemOpen
  \bibfield  {author} {\bibinfo {author} {\bibfnamefont {Y.}~\bibnamefont
  {Enomoto}}, \bibinfo {author} {\bibfnamefont {N.}~\bibnamefont {Kuroda}},
  \bibinfo {author} {\bibfnamefont {K.}~\bibnamefont {Michishio}}, \bibinfo
  {author} {\bibfnamefont {C.}~\bibnamefont {Kim}}, \bibinfo {author}
  {\bibfnamefont {H.}~\bibnamefont {Higaki}}, \bibinfo {author} {\bibfnamefont
  {Y.}~\bibnamefont {Nagata}}, \bibinfo {author} {\bibfnamefont
  {Y.}~\bibnamefont {Kanai}}, \bibinfo {author} {\bibfnamefont
  {H.}~\bibnamefont {Torii}}, \bibinfo {author} {\bibfnamefont
  {M.}~\bibnamefont {Corradini}}, \bibinfo {author} {\bibfnamefont
  {M.}~\bibnamefont {Leali}}, \bibinfo {author} {\bibfnamefont
  {E.}~\bibnamefont {Lodi-Rizzini}}, \bibinfo {author} {\bibfnamefont
  {V.}~\bibnamefont {Mascagna}}, \bibinfo {author} {\bibfnamefont
  {L.}~\bibnamefont {Venturelli}}, \bibinfo {author} {\bibfnamefont
  {N.}~\bibnamefont {Zurlo}}, \bibinfo {author} {\bibfnamefont
  {K.}~\bibnamefont {Fujii}}, \bibinfo {author} {\bibfnamefont
  {M.}~\bibnamefont {Ohtsuka}}, \bibinfo {author} {\bibfnamefont
  {K.}~\bibnamefont {Tanaka}}, \bibinfo {author} {\bibfnamefont
  {H.}~\bibnamefont {Imao}}, \bibinfo {author} {\bibfnamefont {Y.}~\bibnamefont
  {Nagashima}}, \bibinfo {author} {\bibfnamefont {Y.}~\bibnamefont {Matsuda}},
  \bibinfo {author} {\bibfnamefont {B.}~\bibnamefont {Juhasz}}, \bibinfo
  {author} {\bibfnamefont {A.}~\bibnamefont {Mohri}}, \ and\ \bibinfo {author}
  {\bibfnamefont {Y.}~\bibnamefont {Yamazaki}},\ }\href@noop {} {\bibfield
  {journal} {\bibinfo  {journal} {Phys. Rev. Letters}\ }\textbf {\bibinfo
  {volume} {105}},\ \bibinfo {pages} {243401} (\bibinfo {year}
  {2010})}\BibitemShut {NoStop}%
\bibitem [{\citenamefont {Diermaier}(2016)}]{DiermaierThesis}%
  \BibitemOpen
  \bibfield  {author} {\bibinfo {author} {\bibfnamefont {M.}~\bibnamefont
  {Diermaier}},\ }\emph {\bibinfo {title} {Determination of the hydrogen
  ground-state hyperfine splitting in a beam and perspectives for
  antihydrogen}},\ \href@noop {} {\bibinfo {type} {{PhD} dissertation}},\
  \bibinfo  {school} {Technische Universit\"at Wien} (\bibinfo {year}
  {2016})\BibitemShut {NoStop}%
\end{thebibliography}%

\end{document}